\documentclass{emulateapj}

\slugcomment{To appear in ApJ Letters} 
\shorttitle{Quiet Sun internetwork magnetic fields from the 
inversion of {\em Hinode} measurements} 
\shortauthors{Orozco Su\'arez et al.\/}

\begin{document}

\title{Quiet Sun internetwork magnetic fields from the inversion of 
{\em Hinode} measurements}

\author{D.\ Orozco Su\'arez\altaffilmark{1}, L.R.\ Bellot
Rubio\altaffilmark{1}, J.C.\ del Toro Iniesta\altaffilmark{1}, S.\
Tsuneta\altaffilmark{2}, B.W.\ Lites\altaffilmark{3}, K.\ 
Ichimoto\altaffilmark{2}, \\ Y.\ Katsukawa\altaffilmark{2}, S.\
Nagata\altaffilmark{4}, T.\ Shimizu\altaffilmark{5}, R.A.\
Shine\altaffilmark{6}, Y.\ Suematsu\altaffilmark{2}, T.D.\
Tarbell\altaffilmark{6}, A.M.\ Title\altaffilmark{6}}

\altaffiltext{1}{Instituto de Astrof\'{\i}sica de
Andaluc\'{\i}a (CSIC), Apdo.\ de Correos 3004, 18080 Granada, Spain} 

\altaffiltext{2}{National Astronomical Observatory of Japan,
2-21-1 Osawa, Mitaka, Tokyo 181-8588, Japan} 

\altaffiltext{3}{High Altitude
Observatory, NCAR, 3080 Center Green Dr.\ CG-1, Boulder, CO 80301, USA}

\altaffiltext{4}{Hida Observatory, Kyoto University, Takayama, Gifu 506-1314,
Japan} 

\altaffiltext{5}{Institute of Space and Astronautical Science, JAXA,
Sagamihara, Kanagawa 229-8510, Japan} 

\altaffiltext{6}{Lockheed Martin Solar
and Astrophysics Laboratory, Bldg.\ 252, 3251 Hanover St., Palo Alto, CA
94304, USA}

%

\begin{abstract}
We analyze \ion{Fe}{1} 630~nm observations of the quiet Sun at disk center
taken with the spectropolarimeter of the Solar Optical Telescope aboard the
{\em Hinode} satellite. A significant fraction of the scanned area, including
granules, turns out to be covered by magnetic fields. We derive field strength
and inclination probability density functions from a Milne-Eddington inversion
of the observed Stokes profiles. They show that the internetwork consists of 
very inclined, hG fields.  As expected, network areas exhibit a predominance
of kG field concentrations. The high spatial resolution of {\em Hinode}'s
spectropolarimetric measurements brings to an agreement the results obtained
from the analysis of visible and near-infrared lines.
\end{abstract}


%

  \section{Introduction}
  \label{sec:intro}

Most of the studied aimed at determining the distribution of field strengths
in the internetwork (IN) quiet Sun have used polarimetric measurements in the
spectral regions around 630~nm and 1565~nm, but their results do not
agree. The visible \ion{Fe}{1} lines at 630.2~nm indicate a predominance 
of kG fields (S\'anchez Almeida \& Lites 2000; Dom\'{\i}nguez Cerde\~na et al.\
2003; Socas-Navarro \& Lites 2004), whereas the infrared lines at 1565~nm
suggest hG fields (Lin 1995; Lin \& Rimele 1999; Khomenko et al.\ 2003;
Mart\'{\i}nez Gonz\'alez et al.\ 2006a; Dom\'{\i}nguez Cerde\~na et al.\
2006). The distribution of IN field inclinations has only been studied 
by Lites et al.\ (1996) and Khomenko et al.\ (2003). 

Here we analyze \ion{Fe}{1} 630~nm measurements of the quiet Sun taken by the
spectropolarimeter aboard {\em Hinode} at the unprecedented spatial resolution
of 0\farcs32. The observed Stokes spectra are inverted to determine the 
distribution of field strengths and inclinations in the observed region.  Our 
results show that most of the IN fields are weak, opposite to what has been 
found from ground-based measurements of the same lines at 1\arcsec.

%

\section{Observations}
\label{sec:obser}

A quiet solar region of 302\arcsec\/$\times$162\arcsec\/ was observed at disk
center on March 10, 2007 using the spectropolarimeter (SP; Lites et al.\ 2001)
aboard {\em Hinode} (Kosugi et al.\ 2007). The SP records the Stokes spectra
of the \ion{Fe}{1} 630.2~nm lines with a wavelength sampling of
2.15~pm~pixel$^{-1}$ and a scanning step of 0\farcs1476. We used an exposure
time of 4.8~s per slit position, resulting in a noise level of
$1.2\times10^{-3} \, I_{\rm c}$ in Stokes $Q$ and $U$, and $1.1\times10^{-3}
\, I_{\rm c}$ in Stokes $V$.  The data have been corrected for dark current,
flat-field, and instrumental cross-talk as explained by Lites et al.\ (2007c).

The scanned area covers both network and internetwork regions.  The rms
intensity contrast of the granulation is about 7.5\%, which represents the
highest angular resolution ever obtained in spectropolarimetric studies of the
quiet Sun. A visual inspection of the circular and linear polarization maps
reveals a wealth of magnetic signals in the field of view (FOV). We estimate
that 87.1\% and 35.5\% of the image show polarization signals larger than 3
and 4.5 times the noise level, respectively.

%

\section{Inversion strategy}
\label{sec:inver}

To derive the vector magnetic field from the observed Stokes profiles we use a
least-square inversion technique based on Milne-Eddington (ME) atmospheres.
We assume a one-component, laterally homogeneous atmosphere together with
stray/scattered light contamination.  The inversion returns the values of 10
free parameters, including the three components of the magnetic field
(strength, inclination, and azimuth) and the stray light factor, $\alpha$. The
stray light intensity profile is evaluated individually for each pixel by
averaging Stokes $I$ within a box 1\arcsec\/ wide centered on the pixel. We do
not consider broadening of the spectra by macroturbulent velocities. The
inversion is applied to the \ion{Fe}{1} 630.15 and 630.25~nm lines
simultaneously, using a Gaussian of 2.5~pm~FWHM to account for the spectral
resolving power of the SP.

Orozco Su\'arez et al.\ (2007a) demonstrated that this strategy results in
accurate magnetic field inferences for fields above 100~G. The field strengths
derived from ME inversions of \ion{Fe}{1} 630~nm measurements at 0\farcs32
resolution are very similar to those present around optical depth $\tau_5 =
0.01$, with rms uncertainties not larger than 150~G. ME inversions turn out
to be largely independent of the noise and field strength initialization,
provided they are run on pixels showing polarization signals above a
reasonable threshold (Orozco Su\'arez et al.\ 2007b). Here we only analyze
pixels with Stokes $Q$, $U$ or $V$ amplitudes larger than 4.5 times their 
noise levels, in order to exclude profiles which cannot be inverted reliably. 
This threshold corresponds to an apparent flux density\footnote{The apparent 
flux density has been calculated by determining the magnetic parameters 
of a ME atmosphere with vertical fields that produces Stokes $V$ signals 
at the level of the noise. The thermodynamic parameters of the model 
have been fixed to the mean values derived from the Hinode 
measurements.} of 13.4~Mx~cm${^{-2}}$.

\begin{figure}
\centering
\epsscale{1.2}
 \plotone{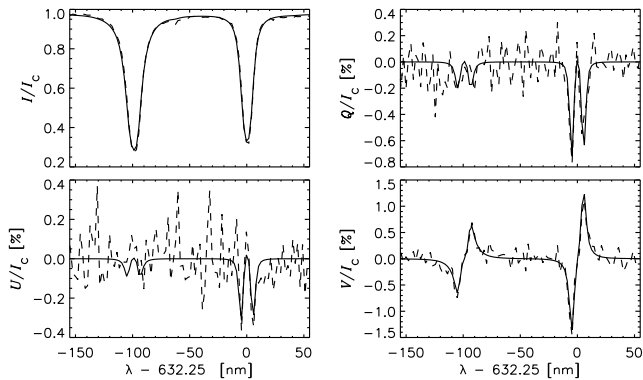}
\caption{Observed ({\em dashed}) and best-fit ({\em solid}) Stokes profiles
emerging from a IN pixel. The field strength and the stray light factor 
are 177~G and 58\%, respectively. \vspace*{.5em}}
\label{fig:fig1}
\end{figure}

Figure~\ref{fig:fig1} shows a sample fit for an individual pixel belonging to
the IN. In this case, the inversion retrieves a field strength of 177~G and a
field inclination of 106$^\circ$. Of interest is that the pixel shows clear
Stokes {\em Q} and {\em U} signals above the noise, just as many other IN
positions. Additional examples can be found in Orozco Su\'arez et al.\ (2007b).

%

\section{Results}
\label{sec:res}

\begin{figure*}
\centering 
\epsscale{1.05}
\vspace*{-0.3em} 
\plotone{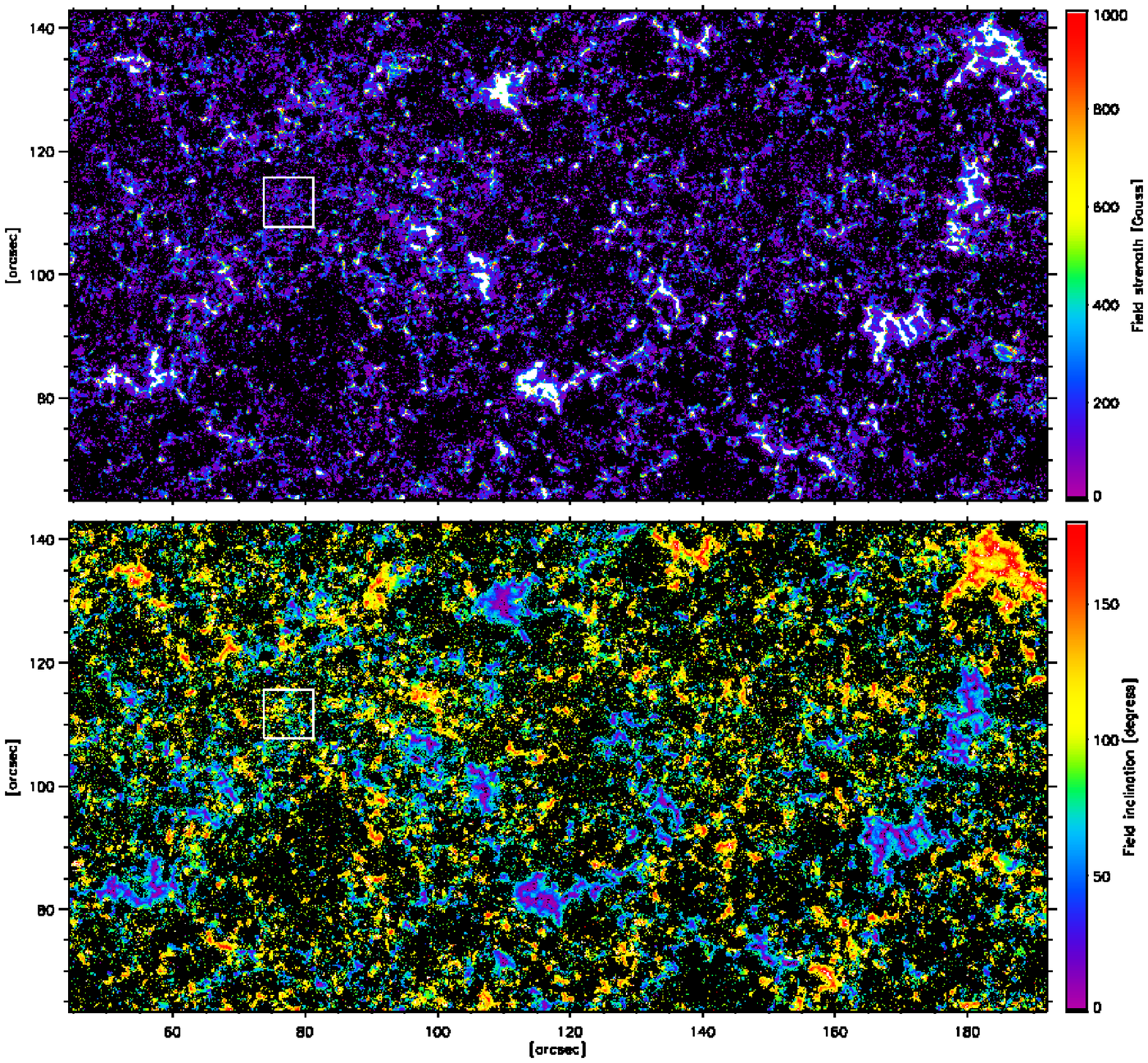}
\epsscale{1.05}
\vspace*{-2em} 
\plotone{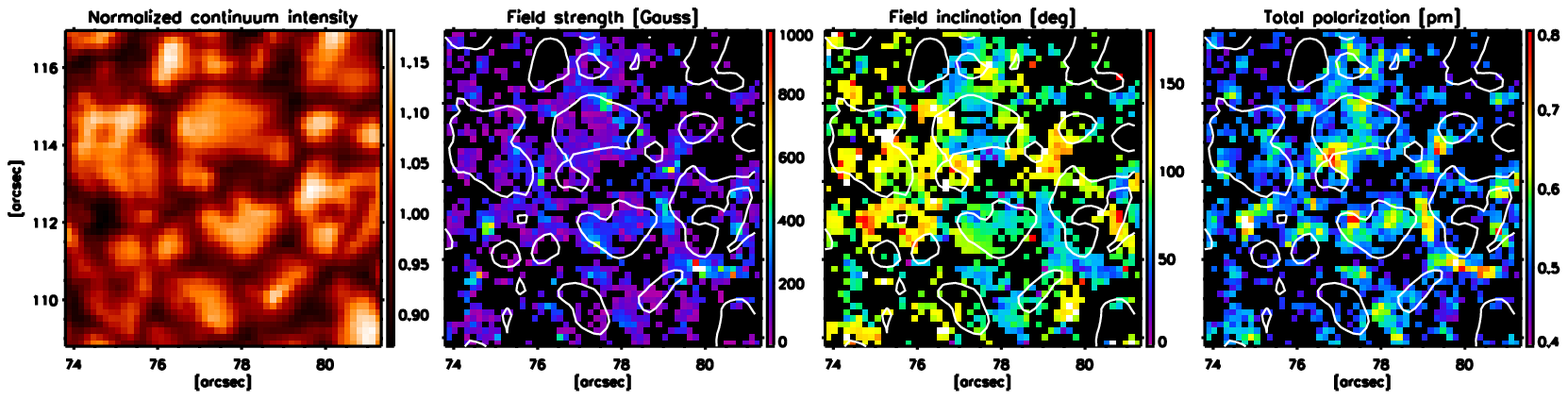}
\caption{Small area of 148\arcsec\/$\times$74\arcsec\ showing the magnetic
field strengths ({\em top}) and inclinations ({\em middle}) inferred from the
inversion. Network and internetwork areas can be easily identified. Black
areas correspond to non-inverted pixels. The field strength color bar has been
clipped at 1000~G ({\em white}). The four bottom panels represent a small IN
area of 7.4\arcsec\/$\times$7.4\arcsec\ (white box in the upper panels). They
display continuum intensities, magnetic field strengths, field inclinations,
and total polarization signals, $\int (Q^2+U^2+V^2)^{1/2}\, {\rm
d}\lambda/I^{\rm QS}_{\mathrm{c}}$. Contour lines represent regions with continuum
intensities $I_{\rm c}/I_{\rm c}^{\rm QS} > 1.05$.}
\label{fig:fig2}
\end{figure*}

Figure~\ref{fig:fig2} shows maps of the retrieved field strength and
inclination for a small portion of the observed area. Black regions represent
pixels which have not been analyzed because of their small signals. In the
field strength map two different regions can be identified: the network,
characterized by strong fields (above 1~kG), and the IN, with much weaker
fields. Supergranular cells are clearly outlined by the network fields. The
inclination map shows that network flux concentrations exhibit nearly 
vertical fields in their interiors and more inclined fields toward the 
edges, suggesting the presence of magnetic canopies. By contrast, IN 
fields are rather horizontal.

The mean unsigned apparent flux density in the FOV is 16.7~Mx~cm$^{-2}$, of
which 9.5~Mx~cm$^{-2}$ correspond to longitudinal flux and 11.3~Mx~cm$^{-2}$
to transverse flux\footnote{The mean flux densities are computed over the FOV
assigning zero fluxes to pixels which were not inverted, so they represent
{\em lower} limits.}.  In IN regions we find a mean flux density of
8.4~Mx~cm$^{-2}$, with 3.4~Mx~cm$^{-2}$ corresponding to longitudinal flux and
7.1~Mx~cm$^{-2}$ to transverse flux. The large occurrence of horizontal fields
confirms the discovery of strong linear polarization signals by Lites et al.\
(2007a,b). The net flux density is 1.7~Mx~cm$^{-2}$ in the full FOV and $-
0.1$~Mx~cm$^{-2}$ in the IN.

The bottom panels of Fig.~\ref{fig:fig2} represent a zoom over a $7\farcs4
\times 7\farcs4$ IN area (white box in the top images) and display continuum
intensities, magnetic field strengths, field inclinations, and total
polarization signals. In the field strength map one can see that most of the
fields are weak. The stronger concentrations are located in intergranular
lanes (the contours outline the granulation). Interestingly, we find
ubiquitous weak fields over granules. The rightmost map just confirms this
finding.  Note also that the fields are more horizontal in granular regions
than elsewhere.

\subsection{Field strength and inclination distributions}

\begin{figure}
\epsscale{1.2}
\centering
\plotone{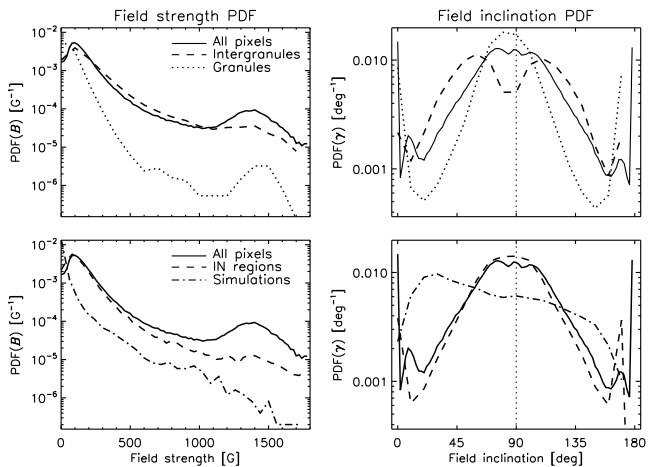}
\caption{Magnetic field strength ({\em left}) and inclination ({\em right})
probability density functions. In the upper panels, the solid, dashed, and
dotted lines stand for all pixels in the FOV, intergranular lanes, and
granules.  In the bottom panels, the solid and dashed lines represent all
pixels in the FOV and IN regions, respectively. Dot-dashed lines show PDFs
from magneto-convection simulations with mean flux density of 10~Mx~cm$^{-2}$.
\vspace*{.5em}}
\label{fig:fig4}
\end{figure}

Figure~\ref{fig:fig4} shows probability density functions (PDFs) for the
magnetic field strength ({\em left}) and field inclination ({\em right}). The
solid lines represent total PDFs considering the $\sim$ 650\,000 pixels
inverted.  The upper panels display PDFs for granules ({\em dotted}) and
intergranular lanes ({\em dashed}). The separation between granular and
intergranular regions has been performed using the continuum intensity and the
inferred line-of-sight velocity.  The bottom panels compare the PDFs of IN
regions ({\em dashed}) with those calculated from the magneto-convection
simulations of V{\"o}gler et al.\ (2005), for a snapshot with mean unsigned
flux of 10~Mx~cm$^{-2}$ ({\em dot-dashed}).  IN areas have been selected 
manually, excluding the boundaries of supergranular cells on purpose.

The peak of the total PDF for the field strength is located at about 90~G. The
curve decreases rapidly toward stronger fields: at around 1~kG it reaches a
minimum and then shows a small hump centered at about 1.4~kG.  Strong fields
($B>$1 kG) are found in only 4.5\% of the pixels, the majority of which
correspond to network areas. It is important to emphasize that the PDF does 
not increase monotonically from 90~G to 0~G. This suggests that the inversions 
are not biased by noise, and that the peak at 90~G is likely solar in origin.
 
The upper left panel of Fig.~\ref{fig:fig4} shows a steeper field strength
distribution in granules as compared with intergranular lanes, i.e., strong
fields are much less abundant in granular regions. Noticeable is the large
fraction of very inclined ($\sim$90$^\circ$) fields in granules. Although
inclined fields are also common in downdrafts, the field lines tend to be more
horizontal over convective upflows (upper right panel of
Fig.~\ref{fig:fig4}). The rapid increase of the PDF near 0$^\circ$ and
180$^\circ$, however, indicates that vertical fields also exist in granules.

In the IN, the field strength distribution reaches a maximum near 90~G and
decreases toward larger fields (bottom left panel of Fig.~\ref{fig:fig4}).
This demonstrates that the IN basically consists of hG flux concentrations.
In the range 1-8~hG, the PDF is well described by a lognormal function $f(B) =
(\pi^{1/2} \sigma B)^{-1} \exp [- (\ln B - \ln B_0)^2/\sigma^2$] with $B_0 =
36.7$~G and $\sigma = 1.2$.  The shapes of the field strength distributions
derived from the simulations and the IN measurements are surprisingly
similar. By contrast, the field inclination PDFs appear to be rather
different, being much flatter in the simulations (Fig.~\ref{fig:fig4}, 
bottom right panel).

\subsection{Stray-light factor distribution}

\begin{figure}[t]
\epsscale{1}
\centering
\vspace*{0.5em}
\plotone{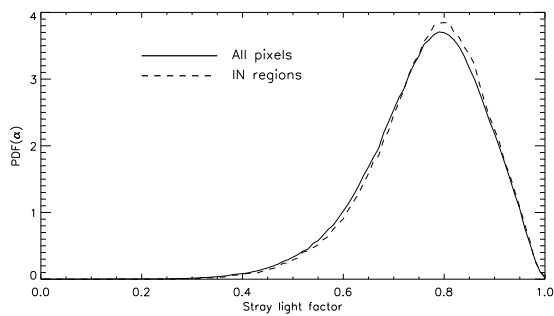}
\caption{PDF of the stray light factor. Solid and dashed lines
represent the full FOV and IN regions, respectively.}
\label{fig:fig5}
\end{figure}

Figure~\ref{fig:fig5} shows the stray-light factor PDF for the full FOV and 
IN regions. Both of them peak at $\alpha \sim 0.8$. For reasons explained in
Orozco Su\'arez et al.\ (2007b), we interpret the stray light contamination as
a degradation of the polarization signals due to diffraction, but it might
also represent filling factors different from 1. In that case, the average
fractional area of the pixel occupied by magnetic fields (given by $1-\alpha$)
would be small, showing a peak at 0.2.

Errors in the stray light determination would immediately lead to different
field strengths and/or inclinations because most of the observed signals are
formed in the weak-field regime. The high-spatial resolution allowed by the
{\em Hinode} SP, however, makes it possible to distinguish between field
strength and stray-light factor unambiguously, even under weak field
conditions. The key ingredient is Stokes $I$: the intensity profile is very
sensitive to small variations in stray-light contamination. Orozco Su\'arez et
al.\ (2007b) have demonstrated that, in practice, the inversion code uses
Stokes $I$ to determine the stray light factor. For more details, including 
an analysis of the $\chi^2$ merit function minimized by the code, the
interested reader is referred to their Sect.\ 6. 



\section{Discussion and conclusions}

The high spatial resolution spectropolarimetric measurements of {\em Hinode}
indicate that most IN fields are weak. This is in agreement with the picture
derived from the more magnetically sensitive \ion{Fe}{1} lines at 1565~nm (Lin
1995; Collados 2001; Khomenko et al.\ 2003; Mart\'{\i}nez Gonz\'alez et al.\
2006b) and from lines showing hyperfine structure such as \ion{Mn}{1} 553~nm
(L\'opez Ariste et al.\ 2006) and \ion{Mn}{1} 1526.2~nm (Asensio Ramos et al.\
2007). Keller et al.\ (1994) also found weak fields in the internetwork using
the \ion{Fe}{1} 525.0~nm lines, although at a lower spatial resolution and
without inclination information.  Our results seem to confirm the mean IN
field strength of $\sim$100~G derived by Trujillo Bueno et al.\ (2004) from a
Hanle-effect interpretation of \ion{Sr}{1} 460.7~nm measurements.

Interestingly, the slope of the field strength distribution in the IN is
similar to that obtained from magneto-convection simulations of comparable
mean flux density. The observed field inclinations, however, turn out to be
significantly larger than those predicted by the simulations. The scenario of
an IN filled by nearly horizontal hG fields is compatible with the large
trasverse magnetic fluxes found in the IN by Lites et al.\ (2007a,b). We still
do not know the origin of such ubiquitous horizontal IN fields, but Lites et
al.\ (2007b) have suggested a number of plausible mechanisms.

In summary, Milne-Eddington inversions of the \ion{Fe}{1} 630~nm lines
observed by {\em Hinode} at 0\farcs32 reveal a predominance of hG fields in
quiet Sun internetwork regions, contrary to what is obtained from the same
lines at 1\arcsec\/. This is the first time that \ion{Fe}{1} 630~nm
observations confirm the weak IN fields indicated by near-infrared
measurements, which may definitely close the discrepancy between the results
derived from both spectral regions.
%

\acknowledgments {\em Hinode} is a Japanese mission developed and launched
by ISAS/JAXA, with NAOJ as domestic partner and NASA and STFC (UK) as
international partners. It is operated by these agencies in
collaboration with ESA and NSC (Norway). This work has been partially
funded by the Spanish Mi\-nisterio de Educaci\'on y Ciencia through
project ESP2006-13030-C06-02.

%

%

\end{document}